\documentclass{webofc}
\usepackage[varg]{txfonts}   

\usepackage{xcolor}

\newcommand{\sqrts}{\sqrt{s_\mathrm{NN}}}

\begin{document}
\title{Directed flow and hyperon polarization at RHIC BES from multi-fluid dynamics}
%
%

\author{\firstname{Iurii} \lastname{Karpenko}\inst{1}\fnsep\thanks{\email{yu.karpenko@gmail.com}} \and
        \firstname{Jakub} \lastname{Cimerman}\inst{1}
}

\institute{Faculty of Nuclear Sciences and Physical Engineering, Czech Technical University in Prague,\\ Břehová 7, Prague, Czech Republic}

\abstract{%
  We present directed flow of protons and pions, as well as mean polarization of $\Lambda$ and $\bar\Lambda$ hyperons computed for Au-Au collisions at $\sqrts=5...19.6$~GeV in MUFFIN model. MUlti Fluid simulation for Fast IoN collisions, or MUFFIN, is a state-of-the-art 3-fluid dynamic model for simulating heavy-ion collisions in the region from a few to a hundred GeV center-of-mass energy. Whereas MUFFIN succeeds to reproduce basic observables in the collision energy range of interest, the slope of the directed flow at mid-rapidity is much steeper as compared to the data, it has unclear EoS dependence and final-state hadronic cascade affects this observable significantly. The excitation function of the $\Lambda$ polarization shows a significant splitting between polarizations of $\Lambda$ and $\bar\Lambda$, which challenges a widespread interpretation that the splitting is affected mainly by the late-stage magnetic field.
}
\maketitle
\section{Introduction}\label{sect-intro}

A goal of heavy-ion collision programs with the center-of-mass energy in the range from a few to to hundered GeV is to investigate the properties of the produced dense baryonic medium, in particular its equation of state (EoS) and transport coefficients. Fluid dynamic approach is instrumental for this goal as it allows to incorporate different equations of state with relative ease.

Fluid dynamic approach has been very successful in its applications to nucleus-nucleus collisions at high energies $\sqrts=200$~GeV and above. There, one typically separates the dynamics into initial state, where the initial hard scatterings are taking place and supposedly lead to isotropisation or effective fluidisation of the medium, and the subsequent fluid stage where the evolution is  governed by fluid dynamical equations.

However, when modelling heavy-ion collisions at the lower energies, one faces a challenge. The Lorentz contraction of the incoming nuclei is not strong, and it takes up to a few fm/c for the two nuclei to completely pass through each other and for all the primary NN scatterings to happen. Dense medium can already be formed in the region where the first nucleon-nucleon scatterings took place, while the last nucleons are still approaching the points of their first interactions.

Multi-fluid dynamics is an elegant though phenomenological way to account for the complex space-time picture of the nucleus-nucleus collision at intermediate energies. In the multi-fluid approach, one approximates the incoming nuclei as two blobs of cold and baryon-rich fluids. A nucleus-nucleus collision is then described as a mutual inter-penetration of the fluids, which slows them down via friction terms. The energy and momentum lost to friction is channeled into creation of the third fluid which represents particles produced in the reaction.

In what follows we present results for directed flow and polarization from the MUlti Fluid simulation for Fast IoN collisions (MUFFIN) model. For a comprehensive description of the model, as well as the reproduction of the basic observables (rapidity disributions and transverse momentum spectra of different sorts of hadrons) we refer the Reader to \cite{Cimerman:2023hjw}.

\section{Model}\label{sect-model}
In the multi-fluid approach, partucularly the 3-fluid one which we employ, the evolution is govened by a coupled set of fluid dynamical equations:
\begin{align}
    \partial_\mu T^{\mu\nu}_\mathrm{p}(x)&=-F_\mathrm{p}^\nu(x)+F_\mathrm{fp}^\nu(x), \nonumber \\
    \partial_\mu T^{\mu\nu}_\mathrm{t}(x)&=-F_\mathrm{t}^\nu(x)+F_\mathrm{ft}^\nu(x), \label{3fh-equations}\\
    \partial_\mu T^{\mu\nu}_\mathrm{f}(x)&=F_\mathrm{p}^\nu(x)+F_\mathrm{t}^\nu(x)-F_\mathrm{fp}^\nu(x)-F_\mathrm{ft}^\nu(x), \nonumber
\end{align}
where the source terms for each fluid are represented via friction terms $F_\mathrm{p,t}^\nu(x)$ and $F_\mathrm{fp,ft}^\nu(x)$. The $F_\mathrm{p}^\nu(x)$ and $F_\mathrm{t}^\nu(x)$ correspond to friction between the projectile and target fluids, acting upon the projectile and target fluids, respectively. The friction terms are defined based on elementary NN scattering, and for more details the Reader is again referred to \cite{Cimerman:2023hjw}. A feature of the coupled fluid dynamical equations (\ref{3fh-equations}) is that the total energy and momentum of all fluids are conserved:
\begin{align*}
    \partial_\mu \left[T^{\mu\nu}_{p}(x) + T^{\mu\nu}_{t}(x) + T^{\mu\nu}_{f}(x) \right]=0.
\end{align*}
Fluid-to-particle transition, aka particlization, is taking place at a hypersurface of fixed effective energy density $\varepsilon_{\rm sw}=0.5$~GeV/fm$^3$. From the segments of this hypersurface, which correspond to fluid freezing out, hadrons are sampled on this hypersurface using Cooper-Frye formula with separate contributions from all fluids. The sampled hadrons are passed on to SMASH hadronic cascade \cite{SMASH:2016zqf} to treat final-state rescatterings and resonance decays.

\section{Results and discussion}\label{sect-results}

The friction terms in MUFFIN have been tuned in order to reproduce pseudorapidity distributions of charged hadrons (measured by PHOBOS), rapidity distribution of net protons (measured by NA49 and BRAHMS) and transverse momentum spectra of protons, pions and kaons (measured by STAR) in heavy-ion (Au-Au or Pb-Pb depending on experiment) collisions at the RHIC BES energy range.

In this proceeding, we present supplementary results from MUFFIN simulations with averaged initial state. We focus on two particular observables, directed flow (more precisely, its slope at mid-rapidity) of protons and pions, and mean (i.e.\ integrated over transverse momentum) polarization of $\Lambda$ hyperons. We have chosen those two observables since they have common prerequisites: partial baryon stopping and finite impact parameter, that also produce finite angular momentum of the created medium.

It has been predicted in early fluid-dynamic calculations that the slope of the directed ﬂow of baryons will turn negative and then positive again as a function of energy if a ﬁrst order phase transition is present in the EoS. More refined studies in modern fluid dynamical model have shown \cite{Steinheimer:2014pfa} that this prediction does not hold. Furthermore, the slope of the directed flow was found to be sensitive not only to the EoS but also to the details of the initial state and particlization prescription.

\begin{figure}[h]
\centering \hspace{-1.2cm}
\includegraphics[width=7cm,clip]{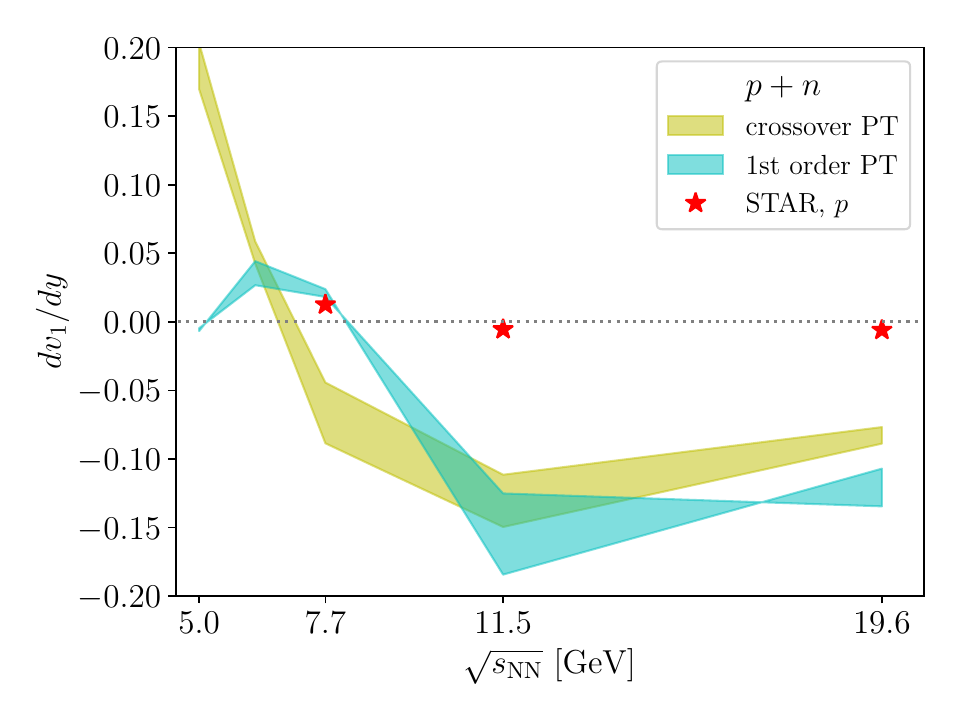}
\includegraphics[width=7cm,clip]{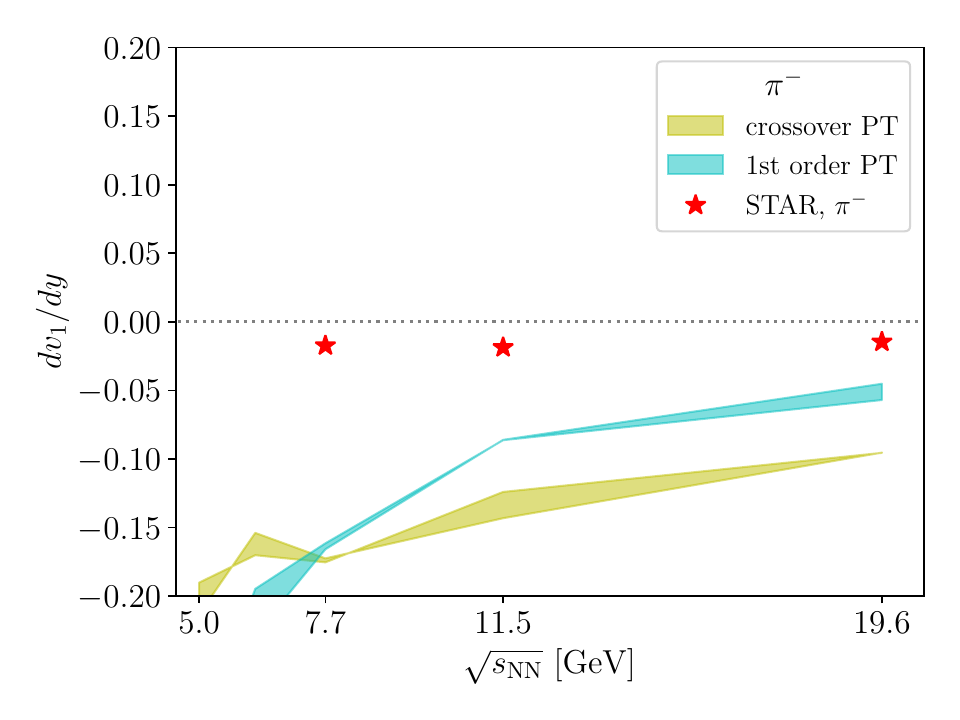}
\caption{Directed flow of nucleons(left) and pions (right), estimated in MUFFIN using reaction-plane method. The simulations are conducted with averaged initial state corresponding to 10-40\% centrality class and an equation of state from chiral model featuring crossover transition to deconfined phase (yellow bands) and bag model EoS featuring first-order phase transition to the deconfined phase (cyan bands). The bands represent a rather simplistic attempt to estimate the uncertainty in the calculation of the slope, stemming from the finite-difference approximation to the derivative. Red stars represent STAR measurements \cite{STAR:2014clz}, and experimental error bars are smaller than the symbol size on this plot.}
\label{fig-v1-eos}       
\end{figure}

\begin{figure}[h]
\centering \hspace{-1.2cm}
\includegraphics[width=7cm,clip]{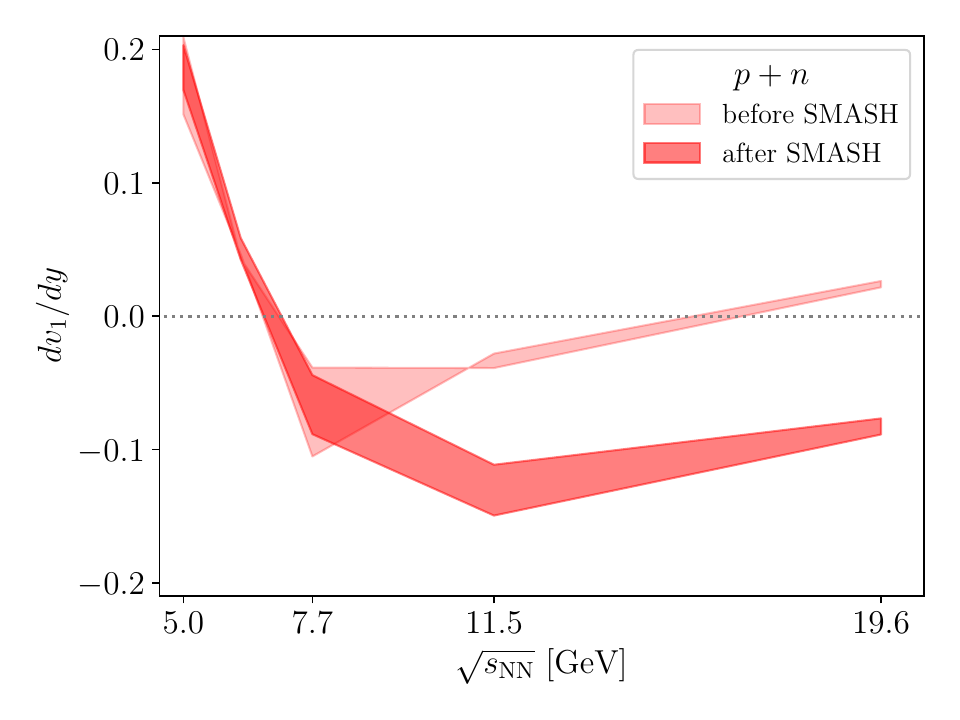}
\includegraphics[width=7cm,clip]{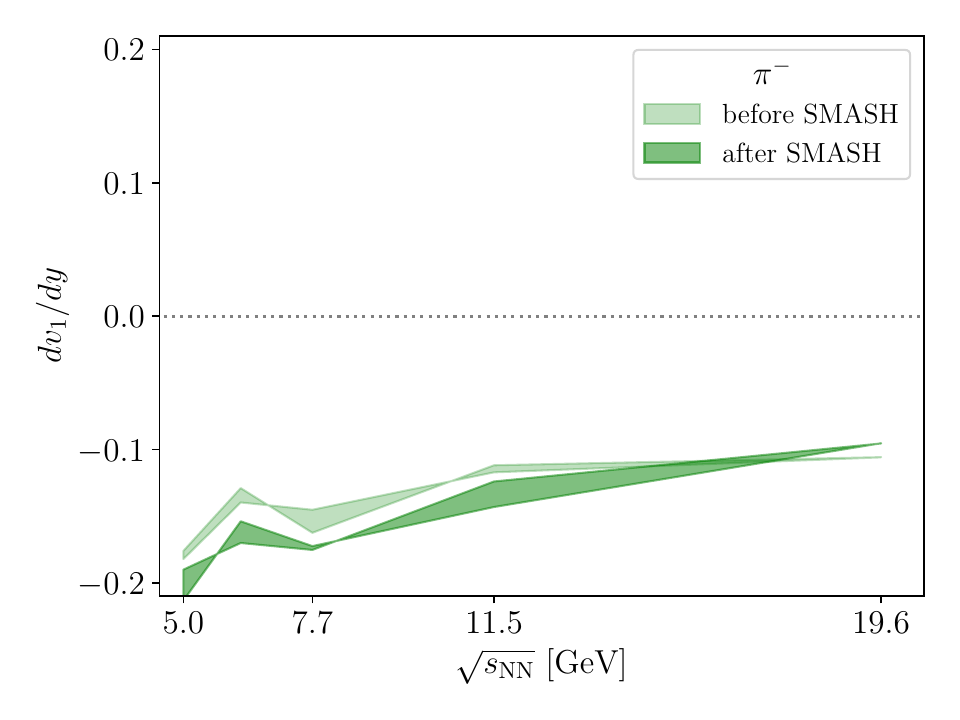}
\caption{Same as Fig.~\ref{fig-v1-eos} but computed before (light-coloured bands) and after (darker-coloured bands) the final-state hadronic cascade.}
\label{fig-v1-smash}       
\end{figure}

This prompted us to examine the excitation function of the slope of directed flow in MUFFIN. In Figure~\ref{fig-v1-eos}, we show the slope of directed flow of the sum of protons and neutrons, as well as negatively charged pions. We sum up protons and neutrons instead of only considering protons, to yield somewhat better statistics for the plots. Here we assume that the effects of electric charge are relatively small as compared to baryon charge in our model. The first impression from the results is that both nucleon and pion $dv_1/dy$ have considerably bigger amplitudes as compared to the data. It is, however, consistent with directed flow reported in other studies \cite{Konchakovski:2014gda}. Swapping the equation of state in the model from chiral EoS with crossover transition to EoS Q with first-order phase transition between the hadronic and QGP phases does not result in a qualitatively different $dv_1/dy$ at $\sqrts>7.7$~GeV. Furthermore, the final-state $dv_1/dy$ of nucleons is significantly affected by the post-hydro phase as one can see from Figure~\ref{fig-v1-smash}. Interestingly, the nucleon $dv_1/dy$ computed right at the particlization (e.g.\ immediately at the end of the fluid stage) has a pronounced non-monotonic collision energy dependence that qualitatively resembles the result from STAR. However, for $\sqrts>7.7$~GeV, the final-state hadronic cascade pulls the $dv_1/dy$ down to large negative values, driving it away from the experimental data points.

\begin{figure}[h]
\centering
\sidecaption
\includegraphics[width=7cm,clip]{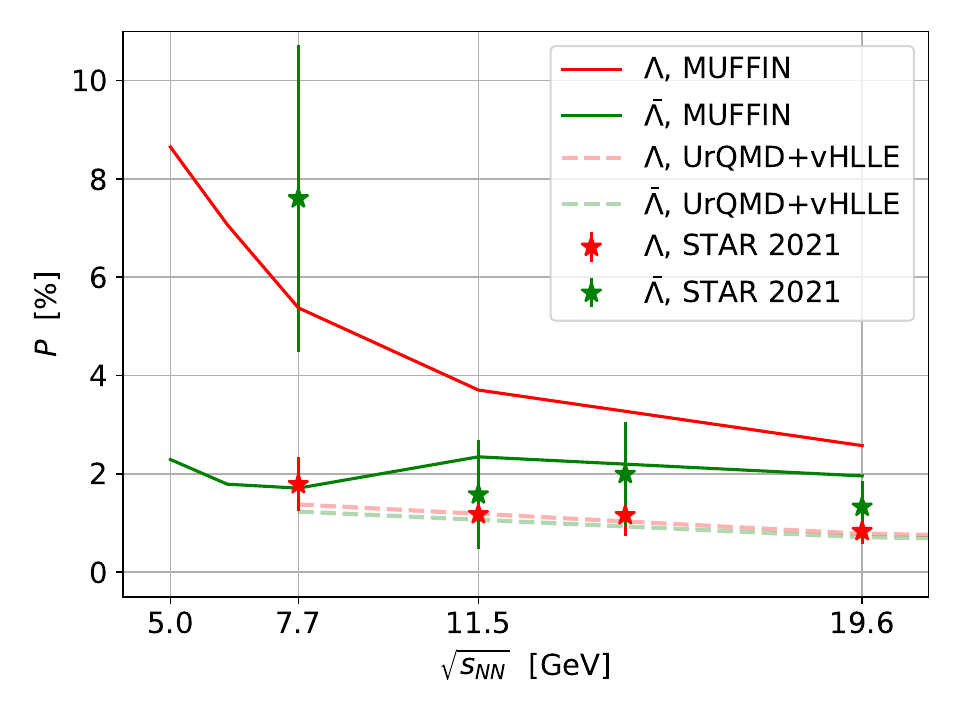}
\caption{Mean polarization of $\Lambda$ and $\bar\Lambda$ hyperons from MUFFIN, computed at particlization. The simulation are performed for 10-40\% centrality class and compared to STAR measurement \cite{STAR:2017ckg} with updated value of $\alpha_{\rm H}$.}
\label{fig-polarization}       
\end{figure}

Next, we examine polarization of $\Lambda$ hyperons. We compute momentum-averaged polarization of $\Lambda$ and $\bar\Lambda$ at mid-rapidity, on the hypersurface of particlization. We add up contributions from thermal vorticity and thermal shear \cite{Becattini:2021iol}. The results are shown on Figure~\ref{fig-polarization}. Note that we compare to STAR data points \cite{STAR:2017ckg}, computed using updated value of $\alpha_{\rm H}$, and to earlier results from a conventional 1-fluid simulation using vHLLE code with UrQMD initial state \cite{Karpenko:2016jyx}. The polarization signal is considerably stronger in MUFFIN as compared to the earlier 1-fluid results. More interestingly, MUFFIN produces a strong splitting between the polarizations of $\Lambda$ and $\bar\Lambda$ at the lower end of RHIC BES energies, which has a sign opposite to the data. This result, however, challenges the widespread interpretation of the splitting, which is attributed to the final-state magnetic field.

Finally, we note that no separate tuning of the model was done for the observables presented in this manuscript.




\begin{thebibliography}{}
%
%

\bibitem{Cimerman:2023hjw}
J.~Cimerman, I.~Karpenko, B.~Tomasik and P.~Huovinen,
Phys. Rev. C \textbf{107}, no.4, 044902 (2023)
[arXiv:2301.11894 [nucl-th]].

\bibitem{SMASH:2016zqf}
J.~Weil \textit{et al.} [SMASH],
Phys. Rev. C \textbf{94}, 054905 (2016)

\bibitem{Steinheimer:2014pfa}
J.~Steinheimer, J.~Auvinen, H.~Petersen, M.~Bleicher and H.~St\"ocker,
Phys. Rev. C \textbf{89}, no.5, 054913 (2014)

\bibitem{STAR:2014clz}
L.~Adamczyk \textit{et al.} [STAR],
Phys. Rev. Lett. \textbf{112}, no.16, 162301 (2014)

\bibitem{Konchakovski:2014gda}
V.~P.~Konchakovski, W.~Cassing, Y.~B.~Ivanov and V.~D.~Toneev,
Phys. Rev. C \textbf{90}, no.1, 014903 (2014)

\bibitem{Becattini:2021iol}
F.~Becattini, M.~Buzzegoli, G.~Inghirami, I.~Karpenko and A.~Palermo,
Phys. Rev. Lett. \textbf{127}, no.27, 272302 (2021)

\bibitem{Karpenko:2016jyx}
I.~Karpenko and F.~Becattini,
Eur. Phys. J. C \textbf{77}, no.4, 213 (2017)

\bibitem{STAR:2017ckg}
L.~Adamczyk \textit{et al.} [STAR],
Nature \textbf{548}, 62-65 (2017)
\end{thebibliography}
\end{document}